\documentclass[conference]{IEEEtran}
\IEEEoverridecommandlockouts

\usepackage{cite}
\usepackage{amsmath,amssymb,amsfonts}
\usepackage{algorithmic}
\usepackage{graphicx}
\usepackage[utf8]{inputenc}
\usepackage{textcomp}
\usepackage{xcolor}
\usepackage{multirow}
\usepackage{float} 
\usepackage{tabularx}
\usepackage{algorithm}
\usepackage{adjustbox}
\usepackage{soul}
\sethlcolor{yellow}

\usepackage{bm}
\makeatletter
\AtBeginDocument{\DeclareMathVersion{bold}
\SetSymbolFont{operators}{bold}{T1}{times}{b}{n}
\SetMathAlphabet{\mathrm}{bold}{T1}{times}{b}{n}
\SetMathAlphabet{\mathit}{bold}{T1}{times}{b}{it}
\SetMathAlphabet{\mathbf}{bold}{T1}{times}{b}{n}
\SetMathAlphabet{\mathtt}{bold}{OT1}{pcr}{b}{n}
\SetSymbolFont{symbols}{bold}{OMS}{cmsy}{b}{n}
\renewcommand\boldmath{\@nomath\boldmath\mathversion{bold}}}
\makeatother
\def\BibTeX{{\rm B\kern-.05em{\sc i\kern-.025em b}\kern-.08em
    T\kern-.1667em\lower.7ex\hbox{E}\kern-.125emX}}

\begin{document}

\title{Network Anomaly Detection for IoT Using Hyperdimensional Computing on NSL-KDD}

\author{\IEEEauthorblockN{\IEEEauthorrefmark{1} Ghazal Ghajari, \IEEEauthorrefmark{1}Ashutosh Ghimire, \IEEEauthorrefmark{2} Elaheh Ghajari, \IEEEauthorrefmark{1} Fathi Amsaad}
\IEEEauthorblockA{\IEEEauthorrefmark{1}Computer Science and Engineering, Wright State University, Ohio, USA }
\IEEEauthorblockA{\IEEEauthorrefmark{2}Computer Science and Engineering, Azad University, Ahvaz, Iran}
Email: \IEEEauthorrefmark{1} \{ghajari.2, ashutosh.ghimire, fathi.amsaad\}@wright.edu
\IEEEauthorrefmark{2} elaheh.ghajari.1@gmail.com
}

\maketitle

\begin{abstract}
With the rapid growth of IoT devices, ensuring robust network security has become a critical challenge. Traditional intrusion detection systems (IDSs) often face limitations in detecting sophisticated attacks within high-dimensional and complex data environments. This paper presents a novel approach to network anomaly detection using hyperdimensional computing (HDC) techniques, specifically applied to the NSL-KDD dataset. The proposed method leverages the efficiency of HDC in processing large-scale data to identify both known and unknown attack patterns. The model achieved an accuracy of 91.55\% on the KDDTrain+ subset, outperforming traditional approaches. These comparative evaluations underscore the model's superior performance, highlighting its potential in advancing anomaly detection for IoT networks and contributing to more secure and intelligent cybersecurity solutions.
\end{abstract}


\section{Introduction}
The Internet of Things (IoT) has emerged as a transformative force across a myriad of industries, enabling unparalleled levels of automation, connectivity, and real-time decision-making. From smart homes and healthcare systems to industrial automation and city infrastructures, IoT devices are pivotal in shaping a more interconnected world. Projections indicate that the number of IoT-connected devices could surpass 500 billion by 2030, reflecting the expansive adoption of these technologies \cite{zikria2021next}. While these advancements offer tremendous opportunities, they also introduce significant security vulnerabilities. IoT networks, characterized by their large-scale, distributed nature and diverse device types, are increasingly being targeted by sophisticated cyberattacks, including distributed denial-of-service (DDoS) attacks and other large-scale service disruptions \cite{mallick2024navigating}. 

IoT environments face unique challenges in ensuring security due to the high volume and velocity of data generated, the heterogeneity of devices, and their often-constrained computational resources. Traditional security measures, such as firewalls and basic intrusion detection systems (IDSs), struggle to provide comprehensive protection. Systems specifically designed for detecting anomalies, known as Anomaly Detection Systems (ADS), have gained prominence for their ability to identify potential threats without relying solely on predefined signatures \cite{diro2024anomaly,ghajari2024hybrid}. However, conventional anomaly detection methods often falter in the dynamic and high-speed IoT landscape, necessitating the development of more adaptive and efficient approaches. 

The rapid evolution of computational methodologies has paved the way for advanced anomaly detection techniques. Artificial intelligence (AI) and machine learning (ML) algorithms, in particular, have demonstrated exceptional promise in automating the identification of abnormal behaviors in IoT data streams. Unlike traditional approaches, these methods can adaptively learn from data patterns and scale to accommodate the growing complexity of IoT networks. Nonetheless, many existing techniques still grapple with challenges related to scalability, efficiency, and accuracy when deployed in real-world IoT scenarios \cite{khattak2024evaluation}.

This study introduces a novel anomaly detection framework tailored for IoT networks, leveraging Hyperdimensional Computing (HDC) combined with a similarity-based classification mechanism. Hyperdimensional Computing is an emerging computational paradigm inspired by the properties of high-dimensional vector spaces, offering advantages in terms of efficiency and robustness. By employing the NSL-KDD dataset—a well-established benchmark for intrusion detection research—the proposed approach systematically addresses the limitations of traditional methods, focusing on the detection of both known and unknown anomalies. The proposed method emphasizes scalability to high-speed IoT environments, adaptability to evolving attack patterns, and computational efficiency suitable for resource-constrained devices.

\section{Related Work}

As the Internet of Things (IoT) continues to expand, the need for effective anomaly detection systems to secure IoT networks has become increasingly critical. Traditional Intrusion Detection Systems (IDSs), which rely on predefined attack signatures, are limited in handling the dynamic, high-dimensional data generated by IoT devices \cite{otoum2021ids}. As a result, there has been significant interest in developing anomaly detection systems (ADS) that can identify novel threats without solely depending on known attack patterns\cite{baweja2020anomaly}.

Machine learning (ML) techniques, such as Random Forests, Support Vector Machines (SVM), and Neural Networks, have been widely applied to detect abnormal behaviors in IoT traffic and various detection systems\cite{pm2024advancements,hashemitaheri2024optical}. For instance, Naive Bayes classifiers and decision tree-based algorithms like J48 have been successfully used to detect intrusions in IoT datasets \cite{razdan2021performance}. However, these models face challenges in high-dimensional, noisy, and unbalanced IoT environments.

Recently, Hyperdimensional Computing (HDC) has emerged as a promising alternative to traditional machine learning techniques \cite{baranpouyan2023hdscc, mohammadi2024hypercell}. HDC mimics the properties of high-dimensional spaces, enabling efficient and noise-resilient encoding of information. This makes HDC particularly suitable for anomaly detection in IoT networks, where large amounts of data from various devices create significant challenges \cite{wang2023hyperdetect}. HDC has been successfully applied to domains like outlier detection and natural language processing\cite{zhang2023adversarial} and has shown potential for intrusion detection by capturing complex patterns in high-dimensional data.

The NSL\textminus KDD dataset, an improved version of the original KDD'99 dataset, has long been the standard benchmark for evaluating intrusion detection models \cite{tavallaee2009detailed}. It has been widely used to assess the performance of various machine learning models, including Autoencoders and SVMs, in network anomaly detection \cite{xu2021improving}. Recent studies have focused on improving the interpretability and accuracy of anomaly detection systems using this dataset by exploring advanced techniques such as deep learning and hybrid models combining multiple algorithms.

Despite significant progress in anomaly detection systems, many existing methods still face challenges related to scalability, adaptability, and real-time detection of novel attacks. The proposed approach in this paper leverages the strengths of Hyperdimensional Computing to overcome these limitations. Using the NSL\textminus KDD dataset, in this study the effectiveness of HDC is demonstrated in identifying both known and unknown attack patterns in IoT networks.

In summary, while traditional and machine learning-based approaches have made significant contributions to anomaly detection in IoT networks, the unique advantages of HDC\textminus such as scalability, efficiency, and robustness to noise—make it a promising approach for future research and practical applications in IoT security.
\begin{figure*}
    \centering
    \includegraphics[width=1\linewidth]{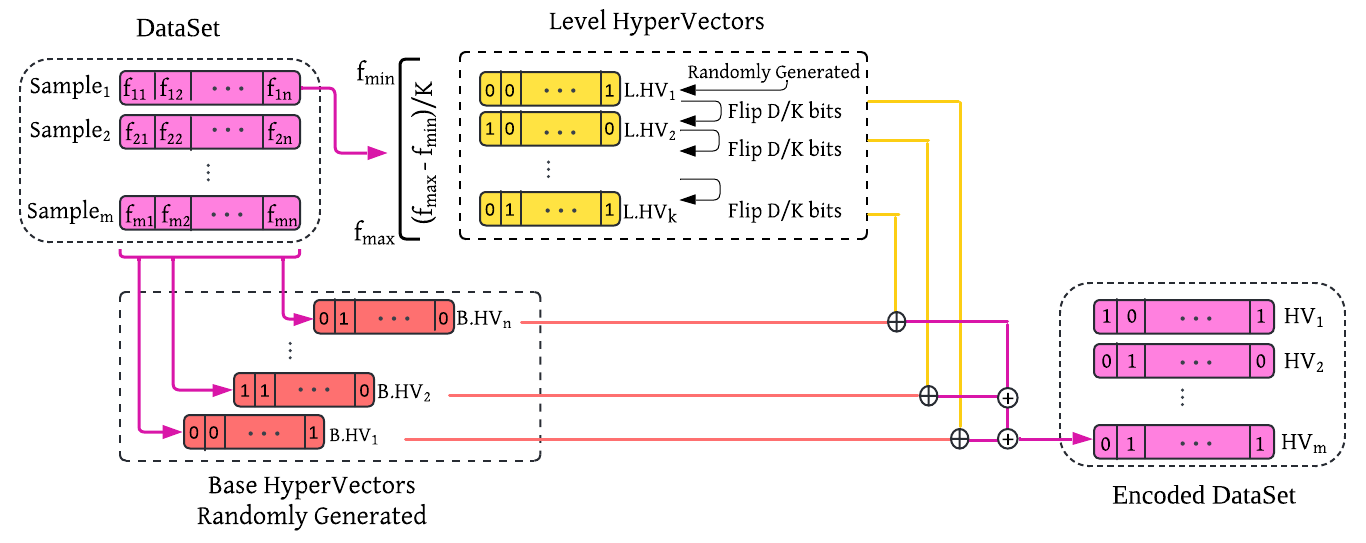}
    \caption{\textbf{Encoding Phase:} dataset features are transformed into high-dimensional binary hypervectors for efficient anomaly detection. This process involves preserving feature identity, quantizing feature values using level hypervectors, and applying XOR operations to form a final representation that captures feature structure and relationships.}
    \label{fig1}
\end{figure*}

\begin{figure}
    \centering
    \includegraphics[width=1\linewidth]{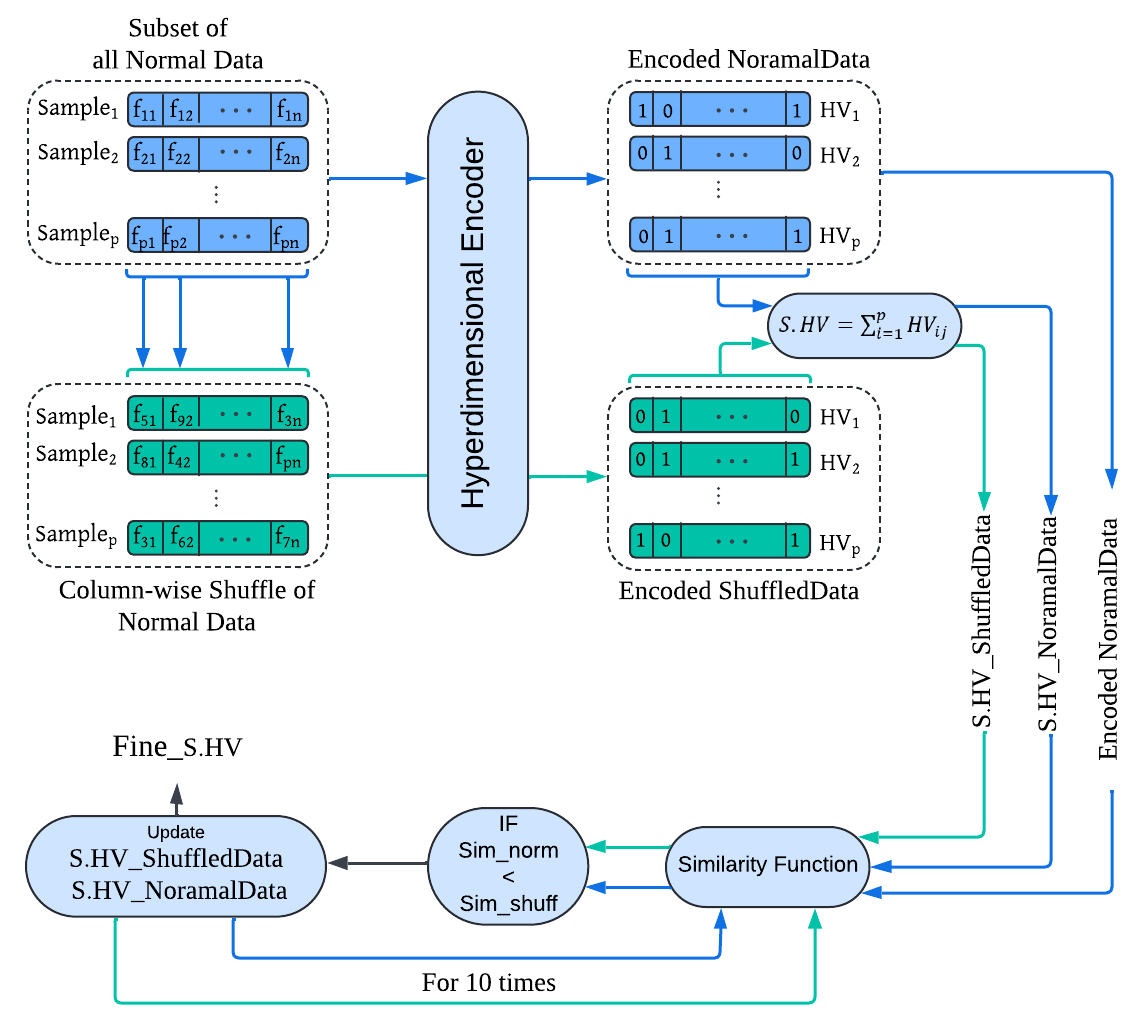}
    \caption{\textbf{Defining one-class hypervector:} Using the hyperdimensional encoder from the previous stage, normal and shuffled datasets are encoded into hypervectors. A one-class similarity vector is iteratively updated using Euclidean similarity, classifying data points as normal or anomalous based on similarity thresholds.}
    \label{fig2}
\end{figure}
\section{Methodology}
The proposed method transforms data into high-dimensional binary hypervectors, preserving its structure and relationships. This efficient approach leverages hyperdimensional spaces for robust anomaly detection. The following sections detail the encoding process and detection strategy.
\subsection{Encoding Phase}
The process of encoding the NSL-KDD dataset, which consists of 125,973 labeled samples and 41 features, into high-dimensional hypervectors involves transforming data points into binary representations while preserving both feature values and their indices. This transformation is crucial for applications in fields such as machine learning and anomaly detection, where high-dimensional spaces provide enhanced representational capacity and robustness. Below is a detailed explanation of the encoding steps, along with the rationale behind each design choice. The data encoding process is depicted in Figure \ref{fig1}.

\subsection* {Step 1: Preserving Feature Identity Using Base Hypervectors}
The first step is to assign each feature in the dataset, represented as $f=<f_1, ... ,f_n>$ , a unique binary hypervector referred to as the Base Hypervector ($B.HV$). These hypervectors are randomly generated binary vectors with a dimensionality of $D = 10,000$.
\begin{itemize}
    \item Reason for High Dimensionality: A high $D$ ensures that the hypervectors for different features are nearly orthogonal, with a Hamming distance close to  $D/2$. This orthogonality minimizes interference between features and preserves their distinct identities.
    \item Random Initialization: Random generation ensures the base hypervectors are uniformly distributed in the high-dimensional space, a property that helps maintain statistical independence between features.
This step ensures that each feature’s unique identity is retained, laying the foundation for encoding feature values in subsequent steps.
\end{itemize}

\subsection* {Step 2: Quantizing Feature Values with Level Hypervectors}
To encode feature values, the range of each feature is divided into  $k$ quantized intervals, and a corresponding Level Hypervector ($L.HV$) is assigned to each interval. This quantization is necessary to map continuous or categorical feature values into discrete levels in high-dimensional space.
\begin{itemize}
    \item Progressive Flipping Scheme:The first level hypervector ($L.HV_1$) is initialized randomly and then For each subsequent interval, $D/k$ random bits are flipped from the previous level hypervector. This ensures that hypervectors for adjacent intervals share a high degree of similarity, while hypervectors for distant intervals become increasingly distinct.
    \item Hamming Distance Control: By flipping a controlled number of bits, the method ensures that $L.HV_1$  and  $L.HV_k$ are nearly orthogonal, with a Hamming distance approximating $D/2$. This property preserves the relationship between feature values, where similar values produce similar hypervectors.
\end{itemize}

\subsection* {Step 3: Mapping Feature Values to Hypervectors}
Once the base and level hypervectors are established, each feature value is mapped to its corresponding quantized interval. This mapping is critical for encoding the relationship between feature indices and their values.
\begin{itemize}
    \item Encoding Operation: The mapping involves an element-wise XOR operation between the feature’s base hypervector ($B.HV$) and the level hypervector ($L.HV$) of its quantized interval. For a feature $f_i$ in the $j^{th}$ interval, the resulting hypervector is:
\begin{center}
    $h_i = B.HV_i \oplus L.HV_j$
\end{center}
\item Preserving Feature Structure: This operation ensures that the encoded hypervector reflects both the feature’s identity (via $B.HV$) and its value (via $L.HV$).
\end{itemize} 
The XOR operation is computationally efficient and maintains the statistical properties of hypervectors, making it a suitable choice for high-dimensional encoding.
\subsection*{Step 4: Aggregating Feature Hypervectors for Data Points}
After generating hypervectors for individual features, the next step is to combine them into a single hypervector representing the entire data point.
\begin{itemize}
    \item Aggregation Method: The hypervectors for all features in a data point are summed element-wise:
    \[H =\sum_{i=1}^{n} h_i = \sum_{i=1}^{n} (B.HV_i \oplus L.HV_i)\]
    \item Interpreting the Result: The summation produces a vector of integer values, where each dimension reflects the cumulative contribution of all features.
\end{itemize}

\subsection*{Step 5: Binarizing the Aggregated Hypervector}
The final step is to convert the aggregated hypervector into a binary format. Binarization is crucial for maintaining the efficiency and robustness of the high-dimensional representation.
\begin{itemize}
    \item Majority Function: For each dimension, the value is compared against a threshold (typically $n/2$, where $n$ is the number of features). The resulting binary value is assigned as follows:
    \[
H{\prime}_i =
\begin{cases}
1 & \text{if } H_i \geq n/2 \\
0 & \text{otherwise}
\end{cases}
\]
\item Threshold Choice: The threshold ensures that the binary representation reflects the dominant contributions across features, reducing noise and enhancing stability.
\end{itemize}

The described methodology systematically transforms datasets into binary hypervectors, preserving both feature indices and values while leveraging the properties of high-dimensional spaces. This encoding scheme is designed for efficiency, robustness, and compatibility with machine learning algorithms. By following this process, the original structure and relationships in the data are effectively captured, enabling powerful analytical capabilities.


\subsection{Define one-class vector and anomaly detection}
In the proposed methodology, a subset of the dataset containing only normal data is selected. This subset undergoes column-wise shuffling, where the data in each column is randomly rearranged, resulting in a synthetic dataset derived from the normal data.

Next, both the normal subset and the shuffled dataset are encoded using the hyperdimensional encoding technique. In the subsequent step, an one\textminus class hypervectors as an initial similarity hypervector ($S.HV$) is defined for the normal ($S.HV_{norm}$) and shuffled ($S.HV_{shuf}$) subsets using a column-wise summation method.
\begin{center}
$S.HV=\sum_{i=1}^pHV_{ij}$    
\end{center}
Where i and j represent the row and column indices, respectively, while $p$ denotes the number of samples in the normal subset and $HV_i$ is the $i-th$ sample in the encoded dataset.
Then, the similarity of the normal subset with the normal similarity hypervector and the shuffled similarity hypervector is calculated using a similarity function and Euclidean normalization.

\begin{center}
$Sim(i)=\frac{\sum_{j=1}^D(HV_i*S.HV)}{ENorm(HV_i)*ENorm(S.HV)}$    
\end{center}

If the similarity of a data point ($HV_i$) to the normal vector is lower than its similarity to the shuffled vector, the data point hypervector ($HV_i$) is added to the normal similarity hepervector and subtracted from the shuffled similarity hepervector.
\begin{center}            
                $S.HV_{norm}+=\alpha*HV_i ~~~  S.HV_{shuf}-=\alpha*HV_i $  
\end{center}
Where $\alpha$ as a learning rate is 0.02.   

This update process for the similarity vectors is repeated 10 times in the proposed model to reach an optimized similarity Hyper vector (Fine\textminus S.HV) to detect normal data. The steps described above are illustrated in Figure \ref{fig2}. 

Finally, based on the obtained similarity measure, the entire dataset is evaluated. Data points with a similarity score higher than a predefined threshold are labeled as normal, while others are labeled as anomalous.
\begin{figure}
    \centering
    \includegraphics[width=1\linewidth]{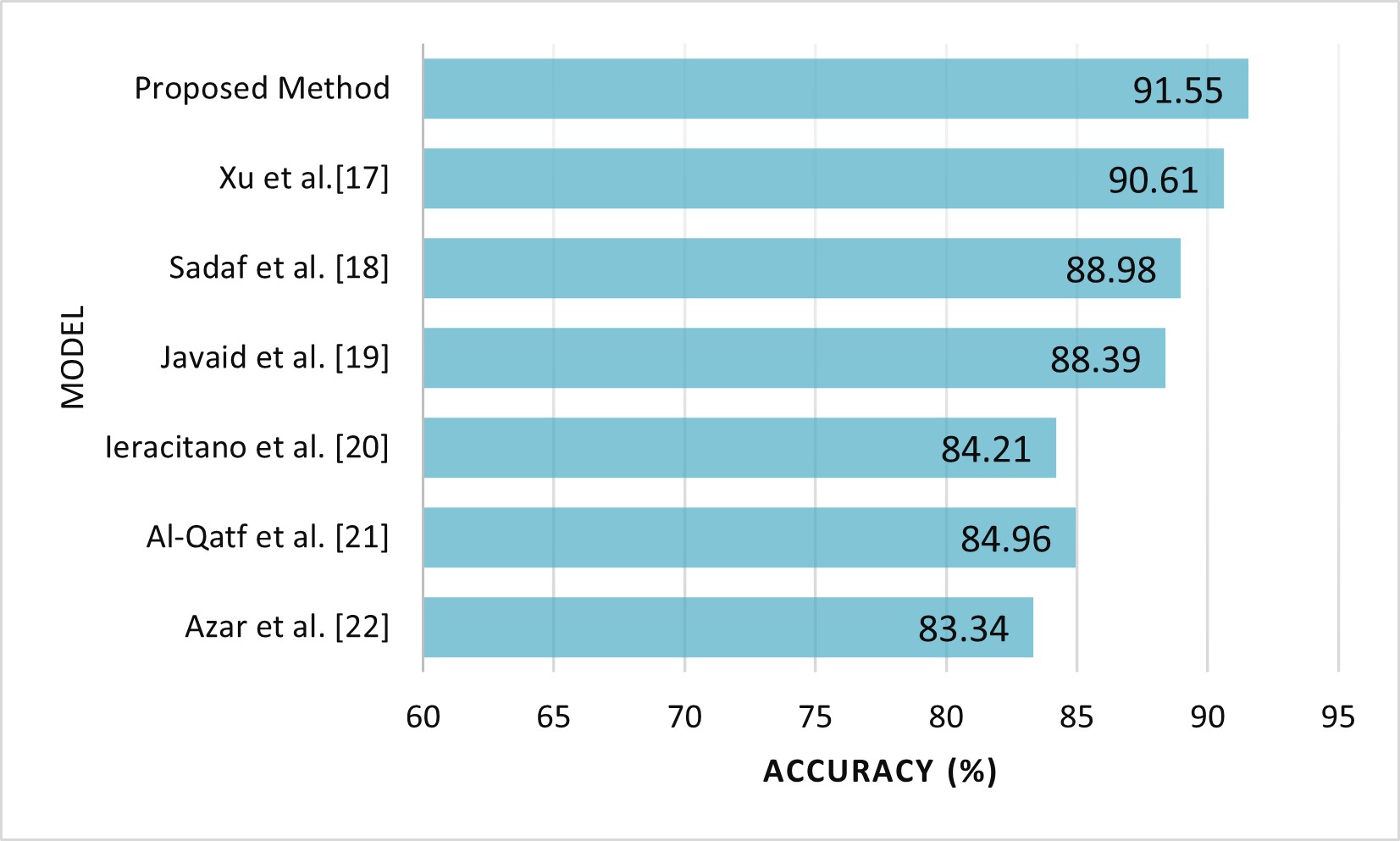}
    \caption{Performance comparison with other approaches on KDDTrain+}
    \label{fig_train}
\end{figure}

\begin{table}[t]
\centering
\caption{Comparison of various models on test+ and test-21.}
\scriptsize 
\resizebox{\columnwidth}{!}{%
\begin{tabular}{p{3cm}p{1.5cm}p{1cm}}
\hline
\multirow{2}{*}{Models} & \multicolumn{2}{c}{Accuracy (\%)} \\ \cline{2-3} 
                        & test+ & test-21 \\ \hline
J48                     & 81.05 & 63.97   \\ 
Naive Bayes             & 76.56 & 55.77   \\ 
NB Tree                 & 82.02 & 66.16   \\ 
Random Forest           & 80.67 & 63.26   \\ 
Random Tree             & 81.59 & 58.51   \\ 
Multi-layer Perceptron  & 77.41 & 57.34   \\ 
SVM                     & 69.52 & 42.29   \\ 
Proposed Method                    & 86.21 & 81.75   \\ \hline
\end{tabular}%
}
\label{tab_test}
\end{table}
\section{Dataset}
The NSL\textminus KDD dataset is an improved version of the KDD’99 dataset, designed to address issues such as redundant records and imbalanced data distribution \cite{tavallaee2009detailed}. By removing duplicate entries, ensuring better data balance, and reducing the dataset's size, NSL-KDD has become an optimized resource for training and evaluating machine learning algorithms. It offers features like smaller size, elimination of unnecessary data, and balanced challenges for learning. The dataset is divided into three subsets: KDDTrain+, KDDTest+, and KDDTest\textminus21, each playing a significant role in testing and evaluating models.

KDDTrain+ is a training dataset with 125,973 records, created by randomly sampling different data groups based on their \#successfulPrediction values. This sampling method results in a more challenging and balanced training set. Similarly, KDDTest+ is an enhanced test dataset with 22,544 records, built by selecting data inversely proportional to the original \#successfulPrediction percentages, providing a better benchmark. KDDTest\textminus21 includes 11,850 records, none correctly classified by all 21 machine learning models in prior studies, designed to test models against difficult, previously misclassified records.

NSL-KDD is essential for IoT anomaly detection, aiding in developing machine learning models to counter threats like DoS attacks and probing. It provides a reliable benchmark for training and evaluating models in dynamic IoT traffic, ensuring effective threat detection even in resource-constrained environments.

Performance evaluations of machine learning models using NSL\textminus KDD demonstrate that the KDDTest+ and KDDTest\textminus21 subsets provide a more realistic assessment of model capabilities. For instance, while KDDTest is unsuitable due to its skewed data distribution, KDDTest+ offers balanced data, and KDDTest\textminus21 introduces greater challenges, allowing for a more thorough evaluation of anomaly detection capabilities. Consequently, NSL\textminus KDD serves as a valuable tool for developing and testing anomaly detection algorithms and related applications in IoT environments.
\section{Result}
In this section, the performance of various models and architectures was evaluated. First, the Autoencoder models introduced in the works \cite{xu2021improving, sadaf2020intrusion, javaid2016deep, ieracitano2020novel, al2018deep, yousefi2017autoencoder} were tested on the KDDTrain+ dataset. These models were assessed to investigate how different architectural choices affect performance. The results of these experiments are shown in Figure \ref{fig_train}.

Various models, including J48 (decision tree) \cite{quinlan2014c4}, Naive Bayes \cite{john2013estimating}, NBTree \cite{kohavi1996scaling}, Random Forest \cite{breiman2001random}, Random Tree \cite{aldous1991continuum}, Multi\textminus Layer Perceptron (MLP) \cite{suter1990multilayer}, and Support Vector Machine (SVM) \cite{chang2011libsvm}, all from the Weka toolkit \cite{weka2008}, were evaluated using the KDDTest+ and KDDTest\textminus 21 datasets. The corresponding results of these evaluations are presented in Table \ref{tab_test}. The proposed model outperforms others in terms of Accuracy, as demonstrated by the experimental results.
\section{Conclusion and Future Work}
In this study, a novel approach to network anomaly detection using hyperdimensional computing (HDC) was proposed and evaluated on the NSL-KDD dataset. The method demonstrated significant advantages in handling high-dimensional data and detecting both known and unknown attack patterns, addressing key limitations of traditional intrusion detection systems (IDSs). Comparative evaluations highlighted its superior performance, emphasizing its potential to strengthen anomaly detection frameworks for IoT networks and advance robust cybersecurity strategies.

For future work, this approach will be extended to more diverse and real-time IoT datasets to ensure scalability and adaptability in dynamic environments. Additionally, exploring hybrid models that integrate HDC with advanced machine learning techniques could further improve detection accuracy and efficiency. Priority will also be given to integrating this method into real-world IoT security architectures and evaluating its performance under varying network conditions for practical applications.
\ifCLASSOPTIONcaptionsoff
  \newpage
\fi
\bibliographystyle{IEEEtran}

\end{document}